\documentclass[12pt]{article}
\usepackage{graphicx}

\usepackage{bm}

\begin{document}

\title{Negative Parity $N^*$ Resonances in an Extended GBE}

\author{\begin{tabular}{c} \normalsize Jun He $^1$  and  Yu-Bing Dong $^{2,1}$ \\
\footnotesize \sl $^1$Institute of High Energy Physics, Chinese
Academy of Sciences, Beijing, P. R. China
\\\footnotesize \sl$^2$Chinese Center of Advanced Science and Technology (World Laboratory)\\\footnotesize \sl
P. O. Box 8730, Beijing 100080, P. R. China \end{tabular}}

\date{\today}

\maketitle
\begin{abstract}
In this paper, we calculate the masses and mixing angles of $L=1$
negative parity $N^*$ resonances in an extended GBE
(Goldstone-boson-exchange model) with harmonic-oscillator wave
functions. By using those mixing angles, we get their photoproduction
amplitudes, and compare them with experimental data and
the results of OPE (one-pion-exchange model), OPsE (only
pseudoscalar meson exchange model), and OGE (one-gluon-exchange
model). We find that the extended GBE gives right internal wave
functions. It is essential to extend GBE to include not only
pseudoscalar meson exchanges but also vector meson exchanges.
\end{abstract}
\normalsize
\section{Introduction}

\ \ \ \ Which is the interaction between quarks mediated by,
gluons or mesons? It is an old problem. In one form or another, it
has been used in a wide variety of models for the last two
decades. In some models, such as, QCD-inspired Isgur-Karl model
\cite{IK} and flux-tube model \cite{flux}, which make success in
explaining the spectra of baryons and mesons and their decay
amplitudes, all spin dependencies are assumed to arise from gluon
exchange. On the other hand, other models, such as cloudy bag
model \cite{bag} and chiral quark model \cite{chiral}, assume the
short-distance force between quarks is mediated by the exchange of
nearly massless Goldstone bosons generated by the spontaneous
breaking of chiral symmetry. It is argued that replacement of
gluon exchange by meson exchange solves some problems of the
quark model for baryons, such as spin-orbit problem, and gives
superior description of baryon spectrum.

However, in 2000 , Isgur published his critique\cite{critique} of
review\cite{GR} by Glozman and Riska in which it is proposed that
baryon spectroscopy can be described by OPE without the standard
OGE forces of Refs. \cite{IK} and \cite{OGE}. In the critique, it
is said that baryon internal wave functions are wrong in OPE.
In fact, in a complex system like the baryon resonance,
predicting the spectrum of the states is not a very stringent test
of a model. The prototypical example is the case of the two
$N^*_{1/2^-}$ states. Among models which perfectly describe the baryon
spectrum, there is still a composition of these states since all
values of $\theta_{{1/2}^-}$ from 0 to $\pi$ correspond to
distinct states. OPE predicts $\theta_{1/2^-}=\pm13^\circ$
and $\theta_{3/2^-}=\pm8^\circ$. Such a $\theta_{1/2^-}$ will have
almost no impact on explaining the anomalously large N$\eta$
branching ratio of the $N^*(1535)_{1/2^-}$ and the anomalously
small N$\eta$ branching ratio of the $N^*(1650)_{1/2^-}$. In our
previous calculation \cite{pre}, by using the mixing angles of
Chizma and Karl \cite{CK}, we find OPE fails to explain the photo- and
electro- production amplitudes for negative parity $N^*$ resonances.

In the argument of Glozman \cite{Gresp}, it's said that only a
$\pi$-exchange tensor force was used for an estimate in
Ref.\cite{GR}. Within GBE picture there are two sources for
tensor force: $\pi$-like exchange and $\rho$-like exchange mechanisms.
For quark separations smaller than $\cong$ 0.6 fm, which is the scale relevant
for baryon structure, the $\rho$-like, $e. g.$ two-pion interaction, is dominant over
the one-pion exchange interaction. And the spin-orbit component of
linear scalar confining interaction in the P-shell multiplets can be
overwhelmed by the spin-orbit component of $\rho$-like exchange\cite{twopion}.
This cancellation was requested by Isgur in his critique paper\cite{critique}.
Both of these exchanges supply a spin-spin force with the same
sign, while their tensor force components have opposite signs.
So it is suggested that addition of $\rho$-like tensor potential
will improve the results\cite{Gresp,twopion}.
So far, this improvement remains to be demonstrated. Recently,
Wagenbrunn $et\ al.$  \cite{EGBE} extended GBE from including only
pseudoscalar mesons\cite{GBE} to including all scalar, pseudoscalar and
vector mesons, and including both spin-spin and tensor components.
The extended GBE allows for an accurate description of all light
and strange baryon spectra. In this paper we want to check whether
the extended GBE can gives right internal wave functions of $L=1$
$N^*$ resonances through an explicit calculation of mixing angles
and phototproduction amplitudes.

In the following section, the eigenstates and eigenvalues of $L=1$
$N^*$ resonances will be given. Then the numerical results of
masses, mixing angles and photoproduction amplitudes will be
shown in section 3. Summary and discussion are given in the last
section.

\section{Eigenstates and eigenvalues of $L=1$ $N^*$ resonances}

\ \ In the harmonic-oscillator model, we assume that the
Hamiltonian of three quark system is of the form \cite{IK}:
\begin{equation} H=\sum_i
m_i+H_1 +H_{hyp}\ ,\end{equation} where
\begin{equation}H_1 =\sum_i
\frac{p_i^2}{2m_i} +\sum_{i<j}V_{conf}^{ij}\ ,\end{equation}and
\begin{equation} V_{conf}^{ij}=\frac{1}{2} K r_{ij}^2 +U(r_{ij})\ .\end{equation}
In Eqs. (1)-(3), $p_i$, $m_i$ are momentum and mass of the quark $i$, and $V_{conf}$ is
a confining potential which is assumed to be a flavor-independent
function of the relative quark separation. In our calculation, we
employ following form for the confinement\cite{EGBE}:

\begin{equation}
V_{conf}^{ij}\ =\ V_0\ +\ C\ r_0\ (1\ -\ e^{-r_{ij}/r_0}\ )\ \sim\ \left\{
\begin{array}{lc}
V_0\ +\ C\ r_{ij}\ ,        &  r_{ij}\ \ll\ r_0\ ,\\
V_0\ +\ C\ r_0\ =\ const\ , &  r_{ij}\ \gg\ r_0\ .
\end{array}
\right.
\end{equation}
It is practically of linear form in the inner region and becomes a
constant outside. The constant $V_0$ is needed to shift the
ground-state energy to the nucleon mass of 939 Mev. $C$ represents
the strength of the liner term for $r_{ij}\gg r_0$. It's necessary
to point out that, when we calculate mixing angles, we don't need
the explicit form of confining potential. This point of view will
be clearly seen later.

$H_{hyp}$ in Eq. (1) is the hyperfine interaction. The extended GBE
assumes pseudoscalar + vector + scalar exchanges. Then chiral interaction
reads\cite{EGBE}:
\begin{eqnarray}H_{hyp}&=&V^{ps}(ij)+V^v(ij)+V^s(ij)\nonumber\\
&=&\sum_{a=1}^3\ [\ V_\pi(ij)\ +\ V_\rho(ij)\ ]\ \lambda_i^a\ \lambda_j^a
+\sum_{a=4}^7\ [\ V_K(ij)\ +\ V_{K^*}(ij)\ ]\ \lambda_i^a\ \lambda_j^a\nonumber\\
&+&[V_\eta(ij)\ +\ V_{\omega_8}(ij)\ ]\ \lambda_i^8\ \lambda_j^8\
+\ \frac{2}{3}\ [V_{\eta'}(ij)\ +\ V_{\omega_0}(ij)\ ]\ +\ V_\sigma(ij)\ ,\end{eqnarray}
with $\lambda^a_i$ the Gell-Mann flavor matrices.
The pseudoscalar meson nonet ($\gamma=\pi,K,\eta,\eta'$) comes with spin-spin and tensor forces
\begin{eqnarray}V_\gamma(ij)\ =\ \ V^{SS}_\gamma(\overrightarrow{r}_{ij})\ \overrightarrow{\sigma}_i
\cdot\overrightarrow{\sigma}_j+\
V^{T}_\gamma(\overrightarrow{r}_{ij})\ [\ 3\
(\hat{\overrightarrow{r}}_{ij}\cdot \overrightarrow{\sigma}_i)\
(\hat{\overrightarrow{r}}_{ij}\cdot
\overrightarrow{\sigma}_j)-\overrightarrow{\sigma}_i
\cdot\overrightarrow{\sigma}_j\ ]\ ,\end{eqnarray} where the
spatial parts have the forms
\begin{eqnarray}V^{SS}_\gamma(\overrightarrow{r}_{ij})\ =\ \frac{g^2_\gamma}{4\pi}\frac{1}{12m_im_j}
[\ \mu^2_\gamma\ \frac{e^{-\mu_\gamma r}}{r}-(\mu^2_\gamma\ +\ \frac{\Lambda_\gamma\ (\ \Lambda^2_\gamma\ -\
\mu^2_\gamma\ )\ r}{2})\frac{e^{-\Lambda_\gamma r}}{r}\ ]\ ,\end{eqnarray}
and
\begin{eqnarray}V^{T}_\gamma(\overrightarrow{r}_{ij})&=&\frac{g^2_\gamma}{4\pi}\frac{1}{12m_im_j}
[\ \mu^2_\gamma\ (1+\frac{3}{\mu_\gamma\ r}+\frac{3}{\mu^2_\gamma\
r^2})\frac{e^{-\mu_\gamma r}}{r} \nonumber\\&&-\ \Lambda^2_\gamma\
(1+\frac{3}{\Lambda_\gamma\ r}+\frac{3}{\Lambda^2_\gamma\
r^2})\frac{e^{-\Lambda_\gamma r}}{r}\ \ -\ \frac{(\
\Lambda^2_\gamma\ -\ \mu^2_\gamma\ )\ (1\ +\ \Lambda_\gamma\ r\
)}{2}\frac{e^{-\Lambda_\gamma r}}{r}]\ .\end{eqnarray} The vector
meson nonet ($\gamma=\rho,K^*,\omega_8,\omega_0$)produces central,
spin-spin, tensor, and spin-orbit components.
\begin{eqnarray}V_\gamma(ij)\ &=&\ V^c_\gamma(\overrightarrow{r}_{ij})\ +\ V^{SS}_\gamma(\overrightarrow{r}_{ij})\ \overrightarrow{\sigma}_i
\cdot\overrightarrow{\sigma}_j\nonumber\\&+&V^{T}_\gamma(\overrightarrow{r}_{ij})\
[\ 3\ (\hat{\overrightarrow{r}}_{ij}\cdot
\overrightarrow{\sigma}_i)\ (\hat{\overrightarrow{r}}_{ij}\cdot
\overrightarrow{\sigma}_j)-\overrightarrow{\sigma}_i
\cdot\overrightarrow{\sigma}_j\ ]\ +\
V_\sigma^{LS}(\overrightarrow{r}_{ij})
\overrightarrow{L}_{ij}\cdot\overrightarrow{S}_{ij}\
.\end{eqnarray} The spin-spin and tensor components of the vector
meson nonet have the same spatial forms as for the pseudoscalar
case above but with the coupling constants replaced in the
following way:
\begin{eqnarray}{\rm spin-spin:}\ \ \ &&\frac{g^2_\gamma}{4\pi}\rightarrow 2\frac{(g^V_\gamma+g^T_\gamma)^2}{4\pi}\ \ \ \ \ \ \ \  \\
{\rm tensor:}\ \ \ \ \ \ \ \ \
&&\frac{g^2_\gamma}{4\pi}\rightarrow
-\frac{(g^V_\gamma+g^T_\gamma)^2}{4\pi} .\end{eqnarray} The scalar
singlet meson ($\gamma=\sigma$) comes with only central and spin-orbit
forces
\begin{eqnarray}V(ij)\ =\ V^C_\gamma(\overrightarrow{r}_{ij})\ +\ V_\gamma^{LS}(\overrightarrow{r}_{ij})
\overrightarrow{L}_{ij}\cdot\overrightarrow{S}_{ij}\ , \ \ \ \ \ \
\ \ \ \  \end{eqnarray} where the central component has the spatial
dependence
\begin{eqnarray}V^{C}_\gamma(\overrightarrow{r}_{ij})\ =\ -\frac{g^2_\gamma}{4\pi}
[\ \frac{e^{-\mu_\gamma r}}{r}-\ (1+\frac{(\
\Lambda_\gamma^2-\mu_\gamma^2)\ r}{2\ \Lambda_\gamma})\
\frac{e^{-\Lambda_\gamma r}}{r}\ ]\ .\end{eqnarray}

All formulae above correspond to a monopole-type
parametrization of the meson-quark interaction vertices, $i.\ e.$
\begin{eqnarray}F(\overrightarrow{q}^2)\ =\ \frac{\Lambda_\gamma^2\ -\ \mu_\gamma^2}
{\Lambda_\gamma^2\ +\ \overrightarrow{q}^2}\ ,\end{eqnarray} where
$\overrightarrow{q}$ is the 3-momentum transferred by the exchanged
meson $\gamma$.

In the equations above, $\overrightarrow{\sigma}_i$ is
Pauli matrix of the quark $i$, and $\mu_\gamma$, $g_\gamma$ are the mass and meson-quark
coupling constant of the exchange meson $\gamma$. The values of Cut-offs $\Lambda_\gamma$
follow the linear scaling laws\cite{EGBE}:
\begin{eqnarray} \Lambda_\gamma\ =\ \Lambda_\pi\ +\ \kappa\ (\mu_\gamma\ -\ \mu_\pi\ )   &&{\rm for\ pesudoscalar\  mesons\ }, \\
 \Lambda_\gamma\ =\ \Lambda_\rho\ +\ \kappa\ (\mu_\gamma\ -\ \mu_\rho\ ) &&{\rm for\ vector\  and\  scalar\  mesons\ }.\end{eqnarray}

In our calculation, we'll leave out all spin-orbit forces and the
central components from the vector meson exchanges as Ref.
\cite{EGBE}. This is further supported on more theoretical grounds by a
study of the two-pion exchange mechanism between constituent quarks in Ref. \cite{twopion}.

We take $U+H_{hyp}$ as perturbation,  then the main part of
Hamiltonian is harmonic potential. In fact, the main part of
Hamiltonian has $SU(6)\otimes O(3)$ symmetry for the spin-flavor
and orbital excitations. This symmetry is broken by
introducing the perturbation including particle exchange
potential. Hence we can say that the admixture origins from symmetry
breaking.

By using the main part of Hamiltonian, we can obtain the standard
wave functions of Isgur-Karl model \cite{IK}:
\begin{eqnarray}S=3/2:&\ \ \ &
\Psi(^4 P)=\frac{1}{\sqrt{2}}\chi^s\{\psi^\lambda \phi^\lambda
+\psi^\rho \phi^\rho\}\ , \\ S=1/2:&\ \ \ \ & \Psi(^2
P)=\frac{1}{2}\{\chi^\lambda \psi^\rho \phi^\rho +\chi^\rho
\psi^\lambda \phi^\rho+\chi^\rho \psi^\rho \phi^\lambda
-\chi^\lambda \psi^\lambda \phi^\lambda\}\ .\end{eqnarray}

The spin angular momentum $S=1/2$, or $3/2$ couples with
the orbital angular momentum $L=1$ to give the total angular
momentum $|L+S|\geq J\geq|L-S|$. As a result there are two states
each at $J=1/2$ and $J=3/2$, namely spin doublet and spin quartet:
$^2P_{1/2}$,$^4P_{1/2}$ and $^2P_{3/2}$,$^4P_{3/2}$. The physical
eigenstates are linear combinations of these two states, and can
be obtained by diagonalizing the Hamiltonian in this space of
states. For example, the $J^P={3/2}^-$states are eigenstates of
the
matrix:\begin{equation}\left(\matrix{\langle^4P_{3/2}|H|^4P_{3/2}\rangle
& \langle^4P_{3/2}|H|^2P_{3/2}\rangle \cr
\langle^2P_{3/2}|H|^4P_{3/2}\rangle
&\langle^2P_{3/2}|H|^2P_{3/2}\rangle}\right)\ .\end{equation} The explicit expressions
of the physical states for the four L=1 negative-parity resonances are:
\begin{eqnarray}|N1700\rangle=cos\theta_{3/2^-}|^4P_{3/2}\rangle+
sin\theta_{3/2^-}|^2P_{3/2}\rangle,
|N1520\rangle=-sin\theta_{3/2^-}|^4P_{3/2}\rangle+
cos\theta_{3/2^-}|^2P_{3/2}\rangle,
\nonumber\\|N1650\rangle=cos\theta_{1/2^-}|^4P_{1/2}\rangle+
sin\theta_{1/2^-}|^2P_{1/2}\rangle,
|N1535\rangle=-sin\theta_{1/2^-}|^4P_{1/2}\rangle+
cos\theta_{1/2^-}|^2P_{1/2}\rangle.\end{eqnarray} where
$\theta_{1/2^-}$, $\theta_{3/2^-}$ are mixing angles, which are
defined as Isgur and Karl \cite{IK77}.

It should be mentioned that matrix elements of spin- and flavor-independent parts of $H$ in
diagonal are equal to each other. And the other two matrix elements in
off-diagonal equal to zero. So mixing angles are independent of
the confining potential and the scalar meson exchange interaction.
\section{Masses and Amplitudes of photoproduction}

\begin{table*}[b]
\caption{Predetermined parameters }
\renewcommand\tabcolsep{0.15cm}\begin{tabular}{ c  c c  c c c c c}  \hline\hline
$m_u=m_d $      &  340MeV  &  $\mu_\pi$                   &   139MeV   &    $\mu_\eta$                &  547MeV     &    $\mu_{\eta'}$    &  958MeV  \\
$\mu_\rho $     &  770MeV  &  $\mu_{\omega_8}$            &   869MeV   &$\mu_{\omega_0}$              &  947MeV     &    $\mu_{\sigma}$   &  680MeV  \\
$g_{ps}^2/4\pi$ &  0.67    &$(g_{v,8}^V+g_{v,8}^T)^2/4\pi$&  1.31      &$(g_{v,8}^V+g_{v,8}^T)^2/4\pi$&    1.31     & $g_{\sigma}^2/4\pi$ &  0.67  \\
$C_0$           &  $2.53$fm$^{-2}$  &  $r_0$  &  $7$fm    & $\alpha$&0.41GeV &&   \\
\hline \hline\end{tabular}
\end{table*}

\ \ \ \ With the formulae above, we first calculate the masses,
$i.\ e.$ physical eigenvalues, of $L=1$ $N^*$ resonances. In our calculation, we adopt
the parameters of Wagenbrunn $et\ al.$\cite{EGBE}. Predetermined
parameters are presented in Table 1.

The free parameters in this work are $\kappa$, $\Lambda_\pi $, and
$\Lambda_\rho$. Here we fix $\kappa$ first, then search the best
fitted masses of the five $L=1$ $N^*$ resonances in the regions of,
$0.4GeV\leq \Lambda_\pi\leq1.4GeV $, and
$0.4GeV\leq\Lambda_\rho\leq1.4GeV $. The constant $V_0$ is
determined by normalizing the ground-state energy to the nucleon
 mass of 939 MeV. We list our results in Table 2.

\begin{table*}[h]
\caption{Best fitted masses of L=1 $N^*$ resonances with different $\kappa$ in the regions of
$0.4GeV\leq\Lambda_\pi\leq1.4GeV $, and
$0.4GeV\leq\Lambda_\rho\leq1.4GeV $. experiment values are from
PDG\cite{PDG}}
\renewcommand\tabcolsep{0.5cm}\begin{tabular}{ c | c c  c c c c }  \hline\hline
$\kappa$  & 0.6     &   0.8  & 1.0     &   1.2    & 1.4 & exp.
\\\hline
$\Lambda_\pi^{fitted}$\ (GeV) &  0.55   &   0.55 &   0.55  &   0.65   &   0.65  &  $-- $    \\
$\Lambda_\rho^{fitted}$\ (GeV)&   1.3   &   1.2  &   1.1   &   0.95   &   0.85  &  $-- $    \\
$V_0         $\ (MeV)         &  -162   &  -165  &  -167   &   -168   &  -169   &  $-- $    \\
$D_{13}(1520)$\ (MeV)         &  1561   &  1558  &  1554   &   1562   &   1561  &  $1515\sim1530$ \\
$D_{13}(1700)$\ (MeV)         &  1705   &  1701  &  1700   &   1704   &   1706  &  $1650\sim1750$ \\
$S_{11}(1535)$\ (MeV)         &  1547   &  1534  &  1527   &   1543   &   1539  &  $1520\sim1555$ \\
$S_{11}(1650)$\ (MeV)         &  1640   &  1644  &  1641   &   1649   &   1647  &  $1640\sim1680$ \\
$D_{15}(1675)$\ (MeV)         &  1659   &  1653  &  1648   &   1662   &   1660  &  $1670\sim1685$ \\

\hline \hline\end{tabular}
\end{table*}
In our calculation, the sum of the confining potential and hyperfine
interaction is smaller than 200MeV, while the main part is about
1900MeV. So perturbation method is reasonable. Though we use the
harmonic-oscillator wave functions, not the wave functions of
Wagenbrunn $et\ al.$ obtained by solving the Schr{\rm \"{o}}dinger
equation with the stochastic variational method, the predicted
masses of five resonances agree with experimental values quite
well.

From the parameters of Table 1, we get the regions of $\Lambda_\pi $, and
$\Lambda_\rho$ for the best mass spectrum. They are  $0.5$GeV$\sim0.8$GeV for
$\Lambda_\pi $, $0.8$GeV$\sim1.3$GeV for $\Lambda_\rho$, respectively. The
fitted values of Wagenbrunn $et\ al.$ are in those regions too.
Thus we can calculate the maximum and minimum of mixing angles of
the extended GBE for photoproduction of the first four $N^*$ resonances
in those regions shown in Table 2. Then by using Eq. (20), we can get the
physical wave functions of those resonances. Since mixing
angles are independent of the confining potential and of the
scalar meson exchange interaction, we don't need the parameters,
$C_0$, $r_0$, $g_{\sigma}^2/4\pi$, and $\mu_{\sigma}$ any longer.

To calculate the electromagnetic transition amplitudes, we use the
electromagnetic interaction of Close and Li \cite{CL} which can be
derived from B-S equation \cite{brodsky}. It avoids the explicit
appearance of the binding potential through the method of McClary
and Byers \cite{MB}. The explicit form is:\begin{eqnarray}
H^{em}&=&\sum_{i=1}^3 H_i=\sum_{i=1}^3 \{-e_i {\bf r}_i \cdot{\bf
E}_ i+i\frac{e_i}{2m^*} ({\bf p}_i\cdot{\bf k}_i{\bf r}_i\cdot
{\bf A}_i+{\bf r}_i\cdot{\bf A}_i {\bf p}_i\cdot{\bf k}_i)-
\mu_i{\bf \sigma}_i\cdot{\bf B}_i\nonumber\\&
-&\frac{1}{2m^*}(2\mu_i-\frac{e_i}{2m^*})\frac{{\bf \sigma}_i}{2}
\cdot[{\bf E}_i\times{\bf p}_i-{\bf p}_i\times {\bf
E}_i]\}\nonumber\\&+& \sum_{i<j}\frac{1}{2M_Tm^*}(\frac{{\bf
\sigma}_i}{2} -\frac{{\bf \sigma}_j}{2})\cdot[e_j{\bf
E}_j\times{\bf p}_i- e_i{\bf E}_i\times{\bf p}_i]\ ,
\end{eqnarray}
where we keep to $ \emph{O}\rm$ $(1/m^2)$, and use long wave
approximation. ${\bf E}_i$ and ${\bf B}_i$ are the electromagnetic
fields, $e_i$, ${\bf \sigma}_i$, $\mu_i$ are the charge, spin, and
magnetic moment of quark $i$. $M_T$ is recoil mass. $m^*_i$ is the
effective quark mass including the effect of long-range scalar
simple harmonic potential, but it is independent on the exchange
potential. So $m_i^*$, or $\mu_i$, in different models can be
treated as the same free parameters.

By insertion of the usual radiation field for the absorption of a
photon into Eq.(21), and by integrating over the baryon
center-of-mass coordinate, we obtain the transverse photoexcited
value over flavor spin and spatial coordinates \cite{C}
\begin{equation}A_\lambda^N =\sum_{i=1}^3 \langle
X;J\lambda|H_i|N;\frac{1}{2}\lambda-1\rangle\ .\end{equation} Here
the initial photon has a momentum $\mathbf{k}||\widehat{z}$. A
simple procedure, that of transforming the wave function to a
basis which has redefined Jacobi coordinates, allows the
calculation of the matrix elements of the $H_1$ and $H_2$
operators to proceed in an exactly similar way to that of the
operator $H_3$. Calculation of the matrix elements of $H_3$ avoids
complicated functions of the relative coordinates in the
''recoil'' exponential.

By using physical wave functions and by using Hamiltonian in Eq.
(21), we calculate the amplitudes of photoexitation from the
ground state N(p,n) to the resonance X by Eq. (22) in
Breit-frame. In the calculation, we follows the convention of
Koniuk and Isgur \cite{KI}. For the photocouplings of the states
made up of light quarks, and the states which are not highly
exited, it should be a reasonable approximation to treat the quark
kinetic mass $m^*_i$ as a constant effective mass, $m^*$. The
recoil mass is kept at $M_T=3m^*$ as Ref. \cite{C}. The origin
of the effect mass of quark here is different from that of
$m_{u,s}$ above. So we adopt $m^*=437$MeV, which is different
from $m_{u,s}=340$MeV. (In fact, the amplitudes aren't sensitive
to the values of $M_T$ and $m^*$.)

A useful measure of the quality of the fit is to
form a $\chi^2$ statistic in the usual way. Introducing a
"theoretical error" \cite{error} avoids overemphasis in the
fitting procedure of a few very well-measured photocouplings.

First, we calculate amplitudes for photoproduction of four states
and $\chi^2$ of those amplitudes at $\kappa=1.2$, which is the
fitted value of $\kappa$ in Ref. \cite{EGBE}. With $\kappa=1.2$,
the scopes of mixing angles in regions of $\Lambda_\pi$$: 0.5$GeV$\sim
0.8$GeV and $ \Lambda_\rho$$: 0.8$GeV$\sim 1.3$GeV are:
\begin{eqnarray} OPsE:&&\ \
22^\circ\leq\theta_{1/2^-}\leq24^\circ,\ \ \ \
-42^\circ\leq\theta_{3/2^-}\leq-35^\circ; \nonumber\\ GBE:&&
-38^\circ\leq\theta_{1/2^-}\leq-23^\circ,\ \ \ \ \
5^\circ\leq\theta_{3/2^-}\leq 8^\circ. \end{eqnarray}In Table 3, we give
results for non-admixture (NA), for OPsE, and for GBE. We also list
the experimental values in last column.

\begin{table*}[t]
\caption{ Breit-frame photoproduction amplitudes at $\kappa=1.2$
using wave functions of no-admixture(NA), of OPsE, and of extended
GBE. Here $\alpha=0.41GeV$, $m^*=0.437GeV$, g=1.3, $M_T=3m^*$.
Amplitudes are in units of $10^{-3}Gev^{1/2}$; a factor of +i is
suppressed for all amplitudes. Experimental values are from PDG
\cite{PDG}}
\renewcommand\tabcolsep{0.1cm}\begin{tabular}{ c  c  c c c c  c cr}  \hline\hline
 state & $A_\lambda^N$ & NA & $\chi^2_{NA}$& OPsE&$\chi^2_{OPsE}$& GBE&$\chi^2_{GBE}$ & Expt. \\\hline
  $N\frac{3}{2}^- (1700)$& $A^p_{\frac{1}{2}}$ & -21 & 0.0&$23\sim29  $    &$3.0\sim3.9$    &$-22\sim-20$   & $0.0\sim0.0$& $ -18\pm13 $    \\
                         & $A^n_{\frac{1}{2}}$ &  19 & 0.1&$21\sim23  $    &$0.2\sim0.2$    &$5  \sim 7 $   & $0.0\sim0.0$& $   0\pm50 $    \\
                         & $A^p_{\frac{3}{2}}$ & -36 & 1.3&$-105\sim-95$   &$9.1\sim11.1$   &$-13\sim -8$   & $0.1\sim0.2$& $  -1\pm24 $    \\
                         & $A^n_{\frac{3}{2}}$ & -14 & 0.1&$40\sim55  $    &$0.8\sim1.4$    &$-51\sim-44$   & $0.7\sim1.0$& $  -3\pm44 $    \\
  $N\frac{3}{2}^-(1520)$ & $A^p_{\frac{1}{2}}$ & -23 & 0.0&$-36\sim-35$    &$0.3\sim0.3$    &$-21\sim-20$   & $0.0\sim0.0$& $ -24\pm9\ $    \\
                         & $A^n_{\frac{1}{2}}$ & -38 & 1.0&$-19\sim-23$    &$2.7\sim3.3$    &$-38\sim-39$   & $0.8\sim0.9$& $ -59\pm9\ $    \\
                         & $A^p_{\frac{3}{2}}$ & 139 & 1.8&$86\sim101$     &$9.9\sim15.1$   &$143\sim144$   & $1.1\sim1.2$& $ 166\pm5\ $    \\
                         & $A^n_{\frac{3}{2}}$ &-125 & 0.4&$-117\sim-110$  &$0.9\sim1.6$    &$-127\sim-128$ & $0.2\sim0.3$& $-139\pm11 $    \\
  $N\frac{1}{2}^-(1650)$ & $A^p_{\frac{1}{2}}$ &  19 & 1.8&$-43\sim-47$    &$14.0\sim15.2$  &$70 \sim100$   & $0.4\sim3.4$& $  53\pm16 $    \\
                         & $A^n_{\frac{1}{2}}$ & -1  & 0.2&$46\sim49$      &$4.4\sim4.9$    &$-49\sim-30$   & $0.3\sim1.4$& $ -15\pm21 $    \\
  $N\frac{1}{2}^-(1535)$ & $A^p_{\frac{1}{2}}$ & 109 & 0.3&$131\sim133$    &$1.3\sim1.4$    &$99\sim 120$   & $0.0\sim0.7$& $  90\pm30 $    \\
                         & $A^n_{\frac{1}{2}}$ & -82 & 1.1&$-91\sim-90$    &$1.7\sim1.8$    &$-83\sim-95$   & $1.2\sim2.1$& $ -46\pm27 $    \\\hline
       $\sum  \chi^2$    &                     &$--$ & 8.1&$--$            &$50\sim59$&$--$           &$7.1\sim9.2$& $--$            \\
\hline \hline\end{tabular}
\end{table*}

In the first two columns of Table 3, the amplitudes without
admixture and $\chi^2$ of those amplitudes are displayed. We can
see that the amplitudes of many states have already agreed with
experimental data well. The other noteworthy information we can
get from the first two columns is that the difference between
$A_{1/2}^{p,n}$ for $N(1650)$ and $A_{1/2}^{p,n}$ for $N(1535)$,
and the difference between $A_{3/2}^{p,n}$ for $N(1700)$ and
$A_{3/2}^{p,n}$ for $N(1520)$, are too large. So the admixture
should not be very large. Otherwise the results which have agreed
with experiment will be destroyed. The third and forth columns in
Table 3 give the results of OPsE, where $\chi^2$ of most amplitudes
increase. $\chi^2$ of $A_{3/2}^{p}$ for $N(1700)$, $A_{1/2}^p$ for
$N(1650)$, or $A_{3/2}^{p}$ for $N(1520)$, is even about 10. All
those amplitudes with large $\chi^2$ are obtained by mixing two amplitudes
 between which there is a large
difference. The sum of $\chi^2$ for twelve amplitudes also increases
from 8.0 to about 55. The fifth and sixth columns present results
of GBE. Almost all amplitudes keep the agreement with experiment.
There is no obvious destruction like OPsE. The sum of
$\chi^2$ keeps at about 8.

To make our conclusion more clear, in Tables 4 and 5, we list the
scopes of mixing angles and sums of $\chi^2$ for four resonances with
OPsE and the extended GBE at $\kappa=0.6,\ 0.8,\ 1.0,\ 1.2,\ 1.4$, respectively,
and compare the results of the extended GBE with OPE and OGE. In the Tables 5, the
mixing angles for OGE and OPE are from Ref. \cite{IK77} and Ref. \cite{CK}, respectively.
"Experimental" values of mixing angles in the last column are from Ref.\cite{expangle}.

\begin{table*}[h]
\caption{Scopes of mixing angles and sums of $\chi^2$ for twelve
amplitudes for photoproduction of four $L=1$ $N^*$ resonances for
OPsE in regions of $\Lambda_\pi$$:0.5$GeV$\sim 0.8$GeV and $
\Lambda_\rho$$:0.8$GeV$\sim 1.3$GeV. Parameters and conventions as
Table 3.}
\renewcommand\tabcolsep{0.43cm}\begin{tabular}{ c | c c  c c c }  \hline\hline
$\kappa$                & 0.6           &   0.8         & 1.0
&1.2 &1.4          \\\hline
$\theta_ {1/2^-}(^\circ)$ &  $23\sim24$   & $23\sim24$    & $23\sim24$    &   $22\sim24$   &   $22\sim23$    \\
$\theta_ {3/2^-}(^\circ)$ & $-44\sim-40$  &  $-44\sim-39$ & $-43\sim-37$  &   $-42\sim-35$ &   $-42\sim-34$     \\
$\chi^2$                & $56\sim62 $   &  $56\sim62 $  & $53\sim60$    &   $50\sim59$   &   $48\sim58$       \\
\hline \hline\end{tabular}
\end{table*}

\begin{table*}[h]
\caption{Scopes of mixing angles and sums of $\chi^2$ for twelve
amplitudes for photoproduction of four $L=1$ $N^*$ resonances for
the extended GBE in regions of $\Lambda_\pi$$:0.5$GeV$\sim 0.8$GeV
and $ \Lambda_\rho$$:0.8$GeV$\sim 1.3$GeV. Parameters and
conventions as Table 3.}
\renewcommand\tabcolsep{0.27cm}\begin{tabular}{ c | c  c c c |c |c|c}  \hline\hline
\multicolumn{5}{c|}{GBE}&\multicolumn{1}{c|}{OPE}&\multicolumn{1}{|c}{OGE}&\multicolumn{1}{|c}{Exp.}\\\hline
$\kappa$        &    0.8        & 1.0           &   1.2     &1.4&
& & \\\hline
$\theta_ {1/2^-}(^\circ)$ &  $-34\sim-15$ & $-36\sim-20$  &   $-38\sim-23$ &   $-39\sim-26$& $25.5$ & $-32$&$-32$     \\
$\theta_ {3/2^-}(^\circ)$ &  $3\sim7  $   & $4\sim7 $     &   $5\sim8$     &   $5\sim8$    & $-52.7$& $6 $ &$ 10$     \\
$\chi^2$                &  $\ 7\sim9 $ & $\ 7\sim9$   &   $\ 7\sim9$  &   $\ 7\sim10$ & $ 80 $ & $11$ & $10$    \\
\hline \hline\end{tabular}
\end{table*}

To analyze our results, we can see that the mixing angles are not sensitive to $\kappa$.
As a result, sum of $\chi^2$ keeps at about 60 for OPsE and at
about 8 for the extended GBE, respectively. The mixing angles of
OPsE which involves all pseudoscalar meson exchanges are similar
to those of OPE. Only $\theta_ {3/2^-}$ has a small improvement. The
sum of $\chi^2$ is still about 60, which is larger than that
of OGE obviously. However, after addition of vector mesons,
 a significant improvement is made. The mixing angles of the extended GBE are
close to those of OGE and "experimental" values. The sum of
$\chi^2$ of the extended GBE remains about 8. With those mixing
angles, N$\eta$ branching ratio can be explained. So we can say
the extended GBE can give right internal wave functions requested
by Isgur.

\section{Summary and discussion}
\ \ \ \ We know that it is hard to judge which model is better
only though spectrum. For example, with the same Hamiltonian but
different wave functions, Ref. \cite{GBE} gave quite different
spectrum. In addition, the hyperfine interaction is quite small compared with
the main part of the mass. Moreover, there are many free parameters and
many kinds of confining potential. In fact, the difference of
confining potentials can be regarded as a hidden free parameter.
The difference of hyperfine interactions may be veiled by adopting
different parameter values. So predicting the spectrum is not
enough to check the small differences between different models,
especially of the hyperfine interactions. However, when we
calculate the mixing angles of states with same $N$ and $L$ as
this paper, the mixing angles are independent of all spin- and
flavor- independent components. We find that the effect of the hyperfine
interaction is dominant. In our calculation, we can see the mixing
angles are more sensitive to hyperfine interaction than the
spectra are. Hence it is necessary to check different models through
mixing angles. It can be regarded as an important way to check the
different hyperfine interactions, especially the tensor part.
Though the extended GBE model yields an improved description of
the light and strange baryon spectra $only$ with $detailed$
hyperfine splittings, the addition of vector mesons is essential
to get good mixing angles and the amplitudes for photoproduction.
After addition of vector mesons, the problem about baryon internal
wave functions of Isgur is solved in this extended GBE.

\Large\begin{flushleft}
{\bf Acknowledgements}\end{flushleft}\normalsize

This work is supported by the National Natural Science Foundation of China
(No. 10075056), by CAS Knowledge Innovation Project No. KC2-SW-N02.

\end{document}